\documentclass[twocolumn]{revtex4-1}

\usepackage{multirow}
\usepackage{amsmath,amsfonts,amssymb,bm}
\usepackage{graphicx}
\usepackage[separate-uncertainty=true]{siunitx}

\usepackage[colorlinks]{hyperref}

\newcommand{\ZJ}[1]{\textcolor{black}{#1}}
\newcommand{\RNum}[1]{\uppercase\expandafter{\romannumeral #1\relax}}

\begin{document}

\title{A high-fidelity heralded quantum squeezing gate}

\author
{Jie Zhao,${}^{1}$ Kui Liu,${}^{2}$ Hao Jeng,${}^{1}$ Mile Gu,${}^{3,4}$ Jayne Thompson,${}^{4}$  Ping Koy Lam${}^{1\ast}$ and Syed M Assad,${}^{1}$ \\
\normalsize{${}^{1}$\it{Centre of excellence for Quantum Computation and Communication Technology,}}\\
\normalsize{\it{Department of Quantum Science, Research School of Physics and Engineering,}}\\
\normalsize{\it{The Australian National University, Canberra ACT 2601, Australia.}}\\
\normalsize{${}^{2}$\it{State key laboratory of quantum optics and quantum optics devices,}}\\
\normalsize{\it{Institute of Opto-Electronics, Collaborative Innovation Center of Extreme Optics,}}\\
\normalsize{\it{Shanxi University, Taiyuan 030006, China.}}\\
\normalsize{${}^3$\it{School of Physical and Mathematical Sciences, Nanyang Technological University, 50 Nanyang Ave, 639798, Singapore.}}\\
\normalsize{${}^3$\it{Complexity Institute, Nanyang Technological University, 50 Nanyang Ave, 639798, Singapore.}}\\
\normalsize{${}^4$\it{Centre for Quantum Technologies, National University of Singapore, Block S15, 3 Science Drive 2, 117543, Singapore.}}\\
}

\date{\today}

\maketitle

\section*{Introductory paragraph}
A universal squeezing gate capable of squeezing arbitrary input states
is essential for continuous-variable quantum
computation~\cite{PRA79062318,PRL112120504}. However, in present
state-of-the-art techniques~\cite{PRA90060302,PRL106240504}, the
fidelity of such gates is ultimately limited by the need to create squeezed vacuum modes of unbounded energy. Here we circumvent this
fundamental limitation by using a heralded squeezing gate. We propose
and experimentally demonstrate a squeezing gate that can achieve near
unit fidelity for coherent input states. In particular, for a
target squeezing of \SI{2.3}{\dB}, we report a fidelity of
\SI{98.5}{\%}.  This result cannot be reproduced by conventional
schemes even if the currently best available squeezing of \SI{15}{\dB}~\cite{PRL117110801} is utilised when benchmarked on
identical detection inefficiencies.  Our technique can be applied to
non-Gaussian states and provides a promising pathway towards
high-fidelity gate operations and fault-tolerant quantum computation.

\section*{main section}

Gaussian operations are essential building blocks for
continuous-variable quantum information processing. With the exception
of the squeezing operation, all other Gaussian operations can be
realised readily with near unit fidelity in quantum optics. However,
current capability to implement squeezing operations with high
fidelity is still limited. This is unfortunate since the squeezing
operation is a prerequisite for universal quantum
computation~\cite{PRA79062318,PRL112120504}. It is also a necessary
component for many other fundamental quantum operations like the
controlled-Z gate~\cite{PRL104250503, PRA71055801,JOSAB2732}, the
quantum non-demolition gate~\cite{PRL101250501}, the control phase
gate~\cite{PRL116180501} and quantum error
correction~\cite{PRL804084}. Moreover, its application to
non-Gaussian states facilitates quantum information tasks
such as decoherence mitigation~\cite{PRL120073603}, preparation of
non-classical states~\cite{PRL113013601} and quantum state
discrimination of coherent-state qubits~\cite{PRA78022320}.

While extensive effort has been devoted to the generation of squeezed
vacuum~\cite{PRL117110801,SQG2016},
the development of a universal squeezing gate that can act on
arbitrary input states has been lagging behind. Recent experiments
have successfully generated a squeezed vacuum with a squeezing
magnitude of \SI{15}{\dB}~\cite{PRL117110801}. In contrast, demonstrations
of a universal squeezing gate have only attained \SI{1.2}{\dB} for a
reliable fidelity of \SI{94}{\%}~\cite{PRA76060301,NatCom28282013}.  High
squeezing levels for vacuum inputs is possible through the parametric
amplification process in an optical cavity. However, extending this
method to arbitrary input states presents significant 
challenges due to cavity loss~\cite{PRL61830,PRL6228,PRL113013601}
and interference effects~\cite{PRL101233602,PRL95233601}.

Instead, current state-of-the-art implementations of a universal
squeezing gate use an ancillary squeezed vacuum as a resource to drive
the squeezing
gate~\cite{NatCom28282013,PRA76060301,PRA90060302,PRL106240504}. Once
the ancillary state has been prepared, the squeezing gate can be
implemented using Gaussian measurements and feed-forward
operations. However, highly squeezed ancilla are required to achieve a
reasonable fidelity. Unit fidelity can only be achieved with an
infinitely squeezed ancilla. Thus in realistic implementations, the
output fidelity will always be limited.

Here, we present and experimentally demonstrate a heralded
squeezing gate that overcomes this limitation. A heralding filter is
implemented in the feed-forward operation whereby an enhancement in
fidelity can be achieved by increasing the filter strength without
requiring more squeezing resources. In contrast to conventional
implementations, this scheme can approach unit fidelity. With
the inclusion of the present squeezing gate, we thus have a complete set
of Gaussian operations that can be implemented with high fidelity.


The universal squeezing gate performs the unitary operation
$S(r_{\mathrm{t}})=\exp\left[ \frac{1}{2}(r_{\mathrm{t}}^* a^2 -
  r_{\mathrm{t}} a^{\dagger 2})\right]$, where $a$ is the annihilation
operator and $r_{\mathrm{t}}$ is the target squeezing strength. Our
squeezing gate is illustrated in Fig.~\ref{fig:1}. First, an optical
parametric amplifier is used to produce an ancillary squeezed vacuum
with squeezing parameter $r_\text{a}$. Next, an input state
$\rho_\text{in}$ is mixed with the ancillary state on a beam splitter
with transmissivity $t_\text{s}$. The reflected mode is then split on
a beam splitter with transmissivity $t_\text{m}$ so that its amplitude
and phase quadratures can be simultaneously measured.  The measurement
outcome, denoted by a complex number $\alpha_{\mathrm{m}}$, is used to
herald a successful squeezing operation by employing a probabilistic
filter. The gate is successfully heralded with probability (see
Methods)
\begin{align}
  \label{eqn:ps}  P_{\mathrm{f}}(\alpha_\text{m})=
  \begin{cases}\exp\left[ \left(1-\frac{1}{g_{\mathrm{f}}}
\right) \left( |\alpha_{\mathrm{m}}|^2{-}
  \alpha_\text{c}^2\right)\right]& ~\text{for}~\left|\alpha_\text{m} \right| < \alpha_\text{c}\,,\\
1& ~\text{for}~ \left|\alpha_\text{m} \right| \geq \alpha_\text{c}\,.
\end{cases}
\end{align}
This filter function, which was previously used to emulate a noiseless
linear amplifier~\cite{PRA86060302,NatPhoton8333,PRA96012319}, depends
on two parameters: the filter strength $g_{\mathrm{f}} \geq 1$, and the
cut-off parameter $\alpha_{\mathrm{c}}$. A large filter strength will
result in a higher output fidelity at the expense of a lower success
probability. When $g_\text{f}=1$, the heralded
squeezing gate reduces to the conventional squeezing gate. The cut-off
parameter determines the operational regime; a
larger cut-off will allow for the squeezing of states with a higher
mean photon numbers. Finally, upon a successful heralding event, the
measurement outcomes of amplitude and phase are rescaled by the
electronic gains $g_{\mathrm{x}}$ and $g_{\mathrm{y}}$, and
fed-forward to the transmitted mode to complete the squeezing
operation. The target squeezing level is determined by the ancillary
squeezing level and the transmissivity of the two beam
splitters (see Supplementary section I).

The faithfulness of a squeezing gate is typically benchmarked by the
fidelity between the output and ideal target state. For Gaussian
inputs within the operational regime, this fidelity is independent of
the input quadrature amplitudes. This is because we operate at the
unity-gain point where the mean quadrature amplitudes of the output
and target coincide (see Supplementary section I).

Figure~\ref{fig:5} illustrates the trade-off between fidelity, target
squeezing, and success probability. We identify two operational
regimes distinguished by the filter
strength. When the filter strength is low, we operate in the first
regime which exhibits a favourable success probability. The majority of
fidelity enhancement can be obtained without dropping below 1\%
success probability. Regardless of the target squeezing, a significant
improvement in fidelity is obtained compared to the conventional
approach. In the second regime, a high filter strength allows near
unit fidelity for any target squeezing
$r_{\mathrm{t}} \leq r_{\mathrm{a}}$, which is impossible
conventionally.  An attractive feature of our scheme is that we can
choose to operate in either regime by simply
tuning the filter strength without reconfiguring the
experimental setup.

We now report the experimental results. An auxiliary squeezed vacuum
with \SI{6.0}{\dB} squeezing and \SI{6.5}{\dB} anti-squeezing was
used as a resource to drive the squeezing gate. In the experiments, we
perform a single quadrature measurement by setting $t_\text{m}
=1$. This allows for a higher success probability compared to a dual
quadrature measurement while maintaining comparable fidelity
enhancement.  In this case, the transmissivity $t_{\mathrm{s}}$ is set
according to $t_{\mathrm{s}}=e^{-2 r_{\mathrm{t}}}$ (see Supplementary section
I). To test the squeezing gate, we prepared several coherent input
states and characterised their
outputs by performing homodyne measurements on the amplitude and phase quadratures. We
implemented at least \num{e6} runs for each input state to generate
enough statistics.

Firstly, we present the results for five input states having different
phases and with magnitudes $|\alpha_\text{in}|$ ranging from \num{0.70} to
\num{1.92} in Fig.~\ref{fig:2}. The target squeezing for these states
vary between \SI{2.3}{\dB} to \SI{10.16}{\dB}. A true
squeezing gate operates on arbitrary inputs irrespective of their
amplitude or phase. This is verified by the measured outputs of our squeezing gate.

Secondly, we characterise the fidelity as a function of target
squeezing in Fig.~\ref{fig:3}a. The best conventional output fidelity
attainable in an idealised experiment using the same ancillary
resource but assuming no loss is plotted as a benchmark. We show that
this benchmark can be surpassed by increasing the filter strength
without requiring a more squeezed ancilla. The trade-off between
fidelity and success probability is illustrated in
Fig.~\ref{fig:3}b. For most runs, the success probabilities are
greater than \num{e-4}.

Thirdly, Fig.~\ref{fig:4}a illustrates the relationship between
fidelity and filter strength. The continuous increase in fidelity as a
function of filter strength agrees with the theoretical model
accounting for experimental imperfections (see Supplementary section~\RNum{4} and Supplementary Fig. 7 for a detailed analysis). The deterministic limit is plotted to identify the minimum
filter strength required to exceed this benchmark. We clearly surpass
this benchmark for all the data sets.

Finally, Fig.~\ref{fig:4}b showcases performance of the heralded
squeezing gate in the high-fidelity regime when the filter strength is
increased to \num{12.63}. For an input magnitude of
$\left|\alpha_\text{in} \right| =2.91$ and target squeezing of
\SI{2.3}{\dB}, we measured a fidelity of \num{0.985 \pm 0.001}. This
fidelity cannot be achieved with the current best squeezed
resource~\cite{PRL117110801} in the conventional scheme subject to the
same homodyne detection efficiency. Assuming idealised experimental
process with zero loss, obtaining this fidelity would require a pure
\SI{10.5}{\dB} squeezed ancilla.

In conclusion, we propose and experimentally demonstrate a heralded
squeezing gate that achieves near unit fidelity for coherent inputs
while requiring only modest ancillary squeezing. Crucially, heralding
circumvents the requirement for a highly squeezed ancilla necessary in
conventional methods. The trade-off between fidelity and success
probability can be tuned at will, and the majority of fidelity
improvement can be achieved without success probability dropping below
1\%. This methodology enables us to synthesise squeezing gates to
a fidelity that would otherwise be impossible for conventional schemes
even with a pure and infinitely squeezed ancillary resource subject to
the same experimental loss. In doing so, our techniques complete
the set of all Gaussian operations that can be experimentally
performed with high fidelity.

There are a number of situations where trading determinism for high
fidelity squeezing can be useful. Notably, in the Supplementary
section II, we illustrate that our techniques can be adapted to squeeze
non-Gaussian states, such as the single-photon state and the
Schr\"{o}dinger's cat state, with near unit fidelity. This provides a
feasible pathway to creating exotic non-classical
states~\cite{PRL91213601} which are key resources for computation and engineering sophisticated nonlinear evolutions~\cite{NatPhy11713,LPR5167}.  High amplitude cat states, for
example, are critical to certain models of universal
continuous-variable computation~\cite{PRA65042313,PRA66023819}. In the cluster state
setting, such non-classical states constitute a resource that enables
a continuous-variable cluster to perform computations that cannot be
efficiently simulated
classically~\cite{PRL97110501,PRA79062318,PRL112120504}. Here, such
resources only need to be prepared offline, and a non-deterministic
mechanism for synthesising them merely adds an overhead to the
preparation procedure. Furthermore, in quantum sensing~\cite{PRA65042313,PRA66023819} and
illumination, the most physically pertinent resource cost is often the
number of photons sent. For example, each probe
risks damaging the sample in biological sensing~\cite{NatPhoton7229}, while in covert sensing, each probe risks
detection by an adversary~\cite{IEEE}. In such scenarios, it becomes
quite reasonable to pay a heavier cost during state preparation to
maximise the efficacy of each probe. Each of these possibilities
merits investigation, whereby one can ascertain the extent
to which it is worthwhile to trade determinism for high fidelity.

\noindent \ZJ{\textbf{Acknowledgements:}} The research is supported by the
Australian Research Council (ARC) under the Centre of Excellence for
Quantum Computation and Communication Technology
(CE110001027). K.L. is supported by the National Natural Science
Foundation of China (Grant No. 11674205; 91536222). M.G. acknowledges
funding from The National Research Foundation of Singapore (NRF
Fellowship Reference No. NRF-NRFF2016-02) and the Singapore Ministry
of Education Tier 1 RG190/17. M.G. thanks the Institute of Advanced
Study at NTU for funding the travel that catalyzed this work. P.K.L is
an ARC Laureate
Fellow.\\

\noindent \textbf{Author contributions:} S.A., J.Z., M.G., J.T., and P.K.L. conceived the experiment. S.A. and J.Z. developed the theoretical
model. J.Z., K.L., H.J., S.A., and P.K.L. planned and performed
the experiment. J.Z. and S.A. analysed the data. J.Z., S.A., H.J., J.T., M.G., and P.K.L. drafted the initial manuscript. All the authors discussed the results and commented on the manuscript. J.Z. and K.L. contribute equally to the work. \\

\noindent \textbf{Competing interests:} The authors declare that they
have no competing financial interests.\\

\noindent \ZJ{\textbf{Additional information:}} \\
\ZJ{\textbf{Correspondence and requests for materials} should be addressed to P.K.L and J.Z.} 

\clearpage
\section*{Materials and Methods}

\begin{figure*}[t]
\begin{center}
\includegraphics[width = 1\columnwidth]{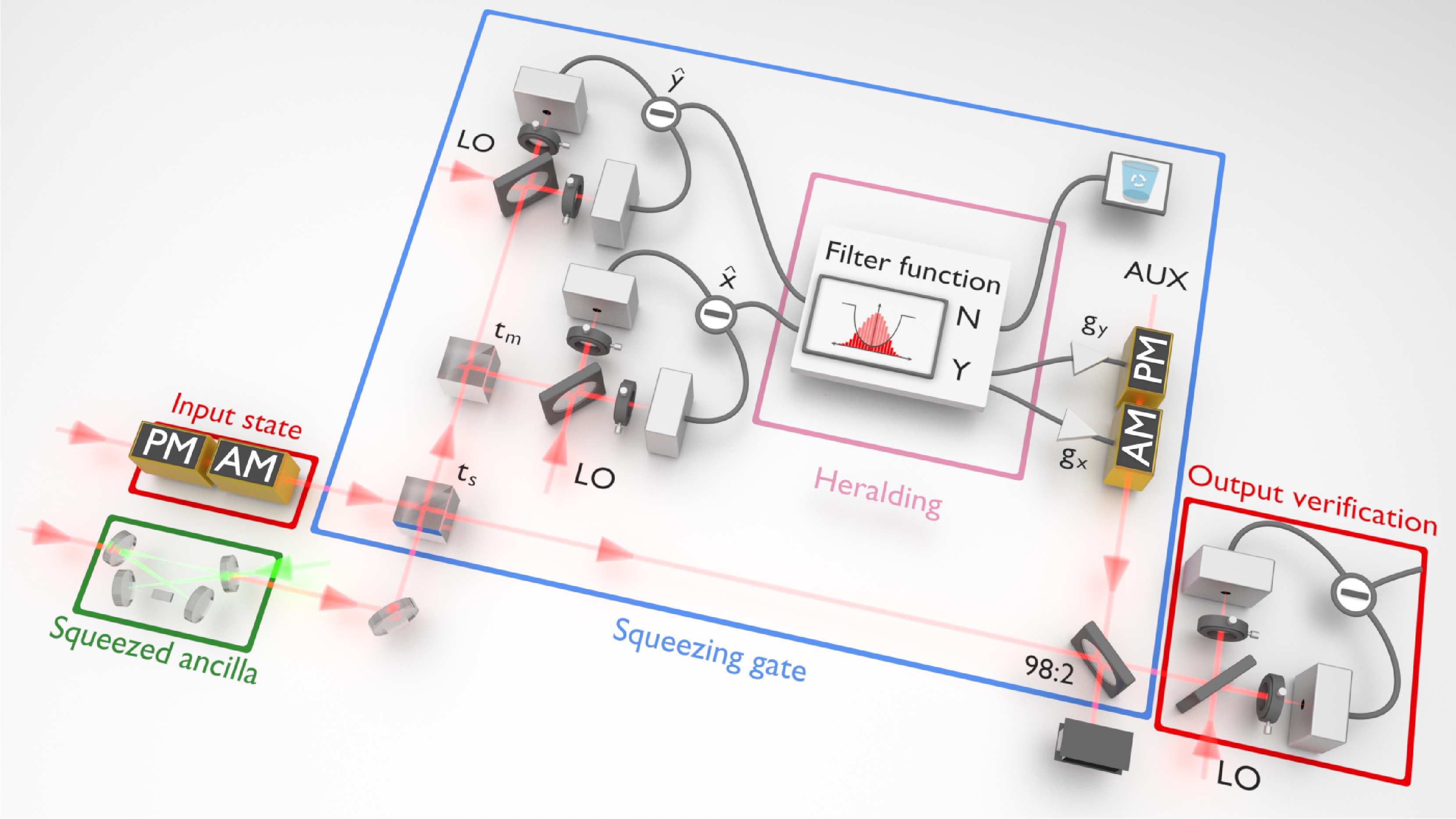}
\caption{\textbf{Experimental layout of the heralded squeezing gate.}
  The gate comprises of three parts. First, the input state and an
  ancillary squeezed vacuum are prepared. Second, the two states are
  mixed on a beam
  splitter with transmissivity of $t_{\mathrm{s}}$. The reflected
  part of the input state is sent to a dual homodyne measurement where
  the two conjugate quadratures, amplitude $\hat{X}$ and phase
  $\hat{Y}$ are measured with a split of $t_{\mathrm{m}}$.
  This
  measurement, in conjunction with a heralded filtering function, feed-forwarding, and a displacement operation
  constitute the core of our probabilistic squeezing gate. Lastly, a verification
  homodyne is employed to characterise the squeezed
  output. The transmissivities $t_{\mathrm{s}}$ and $t_{\mathrm{m}}$
  can be tuned to obtain a trade-off between fidelity and success
    probability. By setting $t_{\mathrm{m}}=1$ in our experimental
    demonstration, we obtain a higher success probability compared to
    the dual homodyne setup without much degradation in the fidelity. AM/PM:
  electro-optic amplitude/phase modulators; LO: local oscillator; AUX:
  auxiliary beam.}
 \label{fig:1}
\end{center}
\end{figure*}

\paragraph*{{\bf Experimental details.}}
As depicted in Fig.~\ref{fig:1}, the experiment consists of four
parts: a squeezed vacuum source, input preparation, squeezing gate
comprised of homodyning and feed-forward, and a homodyne station for
characterising the output state. The main light source is a
continuous-wave frequency-doubled Nd:YAG laser (Innolight Diablo),
producing approximately \SI{300}{\mW} fundamental wave at
\SI{1064}{\nm} and \SI{400}{\mW} second harmonic wave at
\SI{532}{\nm}. The fundamental beam passes through a mode cleaner
cavity with finesse of \num{760} to further purify its spatial mode
and attenuate the high-frequency noise of the laser output. The input
coherent states are created by modulating the fundamental beam at
\SI{4} MHz sideband frequency using a pair of electro-optical
modulators. The squeezed vacuum is prepared in a doubly-resonant
bow-tie cavity where below-threshold optical parametric amplification
(OPA) takes place using a \SI{10.7}{\mm} potassium titanyl phosphate
(KTP) crystal periodically poled with \SI{9}{\um} period. The front
and rear surfaces of the crystal are superpolished and anti-reflection
coated with R\SI{<0.1}{\%} at \SI{1064}{\nm} and R\SI{< 0.2}{\%} at
\SI{532}{\nm}. Three intracavity mirrors are coated to be highly
reflective at both \SI{1064}{\nm} and \SI{532}{\nm} (R\SI{>99.99}{\%}
for the two concave mirrors and R\SI{= 99.85\pm 0.05}{\%} for the flat
mirror), and the input/output coupler has a customised reflection of
\SI{83\pm 1}{\%} at \SI{1064}{\nm} and \SI{73\pm1.2}{\%} at
\SI{532}{\nm}. Up to \SI{11}{\dB} squeezed vacuum can be generated
with a bandwidth of around \SI{36}{\MHz}.

Special care was taken in the implementation of all phase locks
throughout the experiment. The OPA cavity was locked on co-resonance
of both the fundamental beam (\SI{1064}{\nm}) and the pump beam (\SI{532}{\nm}) by
means of the Pound-Drever-Hall technique with a \SI{11.25}{\MHz} phase
modulation on the pump. The same modulation signal was also utilised
to lock the relative phase between the signal beam and the squeezed
ancilla, i.e.\ output of the OPA. The relative phase between the seed
and the pump was carefully controlled using a phase modulation at
\SI{41.5625}{\MHz} on the seed beam. We use this modulation to ensure that
the OPA always operates at the parametric de-amplification, which
yields amplitude-squeezed vacuum. The interference between the seed
and the local oscillators/auxiliary beam on each homodyne station is
controlled similarly with an amplitude modulation (\SI{24.25}{\MHz}) and
phase modulation (\SI{30}{\MHz}) on the signal beam, giving access to the
measurement of an arbitrary quadrature angle.

In the experimental investigation of our squeezing gate, we concerned
ourselves with five different coherent inputs, each being assigned
with a particular target squeezing. By setting $t_{\mathrm{s}}=1$, these input states are characterised by homodyne measurements on amplitude and phase quadratures. The experimental parameters and
measured results for the data depicted in
Figs.~\ref{fig:2}--\ref{fig:4} are tabulated in the Supplementary Table~\RNum{1}.


\paragraph*{{\bf Filter function.}}
The filter function is used to determine if the squeezing gate
operation is successful or not. This is accomplished by picking a
random number from a uniform distribution between 0 and 1 and
comparing that to $P_{\mathrm{f}}(\alpha_\text{m})$ (see
Eq.~\ref{eqn:ps}), where $\alpha_{\mathrm{m}}$ denotes the measurement
outcome of the in-loop dual homodyne detection. The operation is
heralded as successful when the random number is less than
$P_{\mathrm{f}}(\alpha_\text{m})$. In this case, the measurement
outcome is amplified and fed-forward. Otherwise, the operation is
considered to have failed and is aborted. The acceptance rate
$P_{\mathrm{f}}(\alpha_{\mathrm{m}})$ thus 
determines the likelihood of the acceptance of each outcome, which
depends on the amplitude of the input, the filter strength, and the
cut-off. For an initial Gaussian distribution of
$\alpha_{\mathrm{m}}$, the resultant distribution remains Gaussian
provided that the cut-off is sufficiently large, but with its mean
and variance both being amplified by $g_{\mathrm{f}}$
\cite{PRA96012319}. To be concrete, by applying the filter function
upon an unnormalised Gaussian ensemble with mean $\alpha_0$
\begin{align}
\label{eq:filterin}
e^{-|\alpha_\text{m}-\alpha_0|^2}\;,
\end{align}
the ensemble evolves as per multiplying Eq.~\ref{eq:filterin} by the
acceptance rate $P_{\mathrm{f}}(\alpha_{\mathrm{m}})$. \ZJ{The output distribution is essentially a concatenation of two Gaussian distributions joined at the circle $|\alpha| = \alpha_{\mathrm{c}}$. The inputs with $\alpha_{\mathrm{m}}>\alpha_{\mathrm{c}}$ are unaffected, whereas those with $\alpha_{\mathrm{m}}<\alpha_{\mathrm{c}}$ are filtered and become proportional to:} 
\begin{align}
  &e^{-|\alpha_\text{m}-\alpha_0|^2}e^{\left( 1- 1/g_{\mathrm{f}} \right)\left(|\alpha_\text{m}|^2- \alpha_{\mathrm{c}}^2\right)}\nonumber\\
  =&\frac{e^{(g_{\mathrm{f}}-1)|\alpha_0|^2}}{e^{(1-1/g_{\mathrm{f}})\alpha_{\mathrm{c}}^2}}\exp\left( -\frac{|\alpha_\text{m}-g_{\mathrm{f}}\,\alpha_0|^2}{g_{\mathrm{f}}}\right)\;.
\label{eq:filterout}
\end{align}
\ZJ{Note that only the second part ({\it cf.} Eq.~\ref{eq:filterout}) that undergoes post-selection is desired. Therefore, a sufficient cut-off should be able to embrace this Gaussian as much as possible. To be more explicit, it was proposed in~\cite{PRA96012319} that,
\begin{align}
 \label{eq:suffcutoff}
 \alpha_{\mathrm{c}} = g_{\mathrm{f}}^2 |\alpha_{\mathrm{m}}| +\beta g_{\mathrm{f}} \sigma_{\alpha_{\mathrm{m}}} / \sqrt{2} .\
\end{align}
Here $\sigma_{\alpha_{\mathrm{m}}}$ is the standard deviation of the input distribution; $\beta$ quantifies how well the cut-off circle embraces the output distribution. 
}
Upon restoring the proper normalisation of \ZJ{the output distribution}, we obtain the success
probability of the filtering operation as
\begin{align}
 \label{eq:probNLA}
P_{\mathrm{s}} &= \frac{e^{(g_{\mathrm{f}}-1)|\alpha_0|^2}}{\pi\,e^{(1-1/g_{\mathrm{f}})\alpha_{\mathrm{c}}^2}} \iint\limits_{\left|\alpha_\text{m}\right|< \alpha_{\mathrm{c}}} \exp\left(- \frac{\left|\alpha_\text{m}- g_{\mathrm{f}} \alpha_0
  \right|^2}{g_{\mathrm{f}}} \right)  \text{d}^2\alpha_\text{m} \nonumber \\
  &+ \frac{1}{\pi} \iint\limits_{\left|\alpha_\text{m} \right|\geq\alpha_{\mathrm{c}}} \exp\left(-\left|
      \alpha_\text{m}- \alpha_0 \right|^2 \right) \text{d}^2\alpha_\text{m}\;.
\end{align}
We note that a larger cut-off enables a wider operational range of the
filter function, however, at the expense of a decreased success probability. For an input ensemble with a large amplitude, a sufficiently
large $\alpha_{\mathrm{c}}$ is required; otherwise, the part of the
output distribution beyond $\alpha_{\mathrm{c}}$ is subject to
distortion~\cite{PRA96012319}. Hence, $\alpha_\text{c}$ needs to be
carefully tailored according to the input ensemble to ensure a
faithful squeezing operation while still maintaining a reasonable
probability of success~(see Supplementary section \RNum{5}).

\noindent \ZJ{\textbf{Data availability:} The data that support the plots within this paper and other findings of this study are available from the corresponding author upon reasonable request.}

\pagebreak
 \onecolumngrid

\clearpage

\begin{figure}[h]
  \centering
\includegraphics[width = 1\columnwidth]{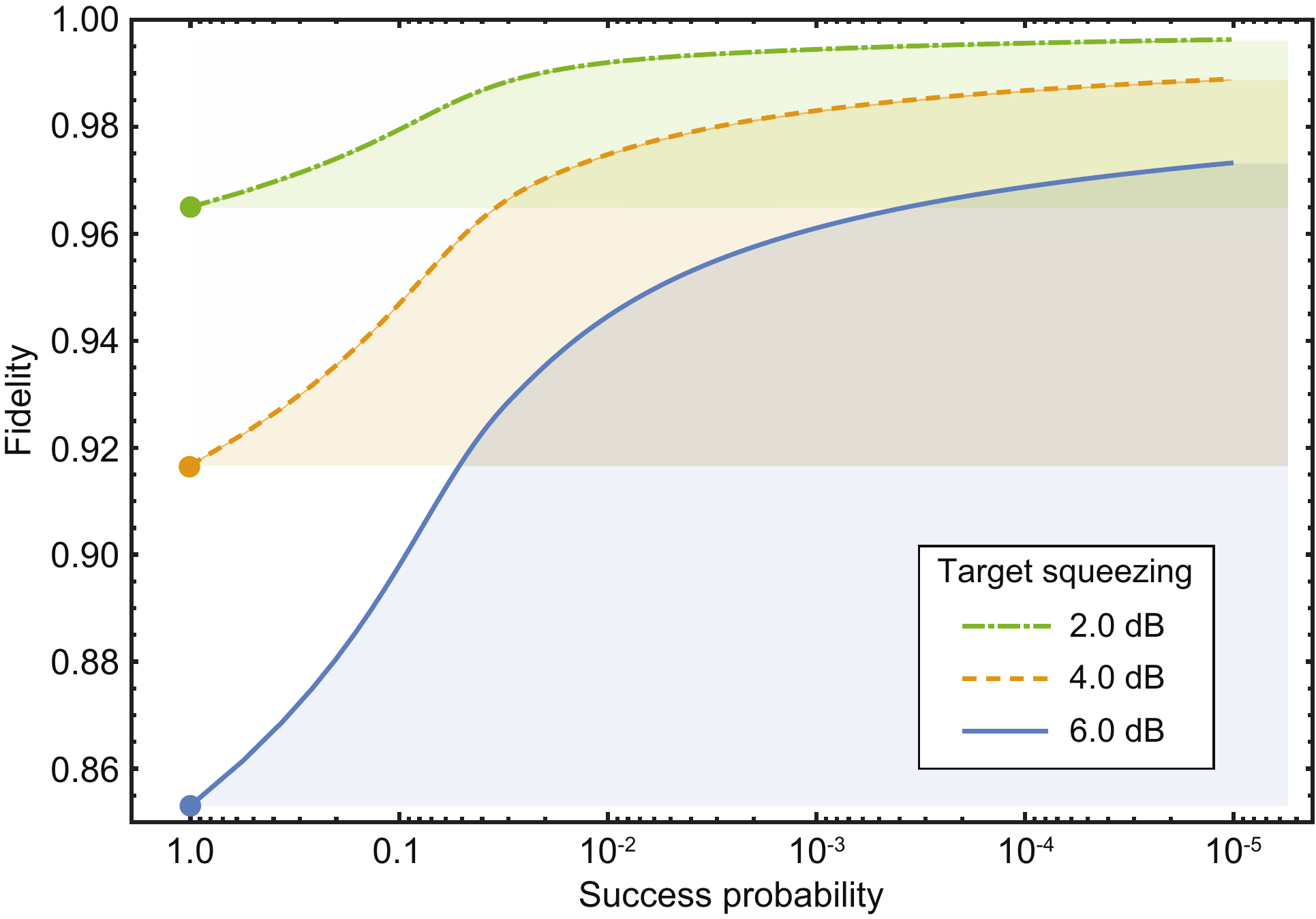}
\caption{\textbf{Fidelity against success probability for various
    target squeezing.} Starting with a pure squeezed ancilla
  of \SI{6}{\dB}, the fidelity is plotted as a function of success probability for
  three values of the target squeezing, being \SI{2}{\dB}, \SI{4}{\dB} and \SI{6}{\dB}.
  For comparison, the fidelity of a conventional
  deterministic squeezing gate is superimposed (dots). In all
  cases, a substantial enhancement in fidelity is achieved with the heralded squeezing gate, at the expense of a lower success probability. The cut-off $\alpha_{c}$ is chosen to include more than \SI{98}{\%} of the
  total statistics to ensure the Gaussianity of the output is preserved~(see Supplementary Materials \RNum{4}).  }
 \label{fig:5}
\end{figure}

\begin{figure}[h]
  \centering
  \includegraphics[width = 1\columnwidth]{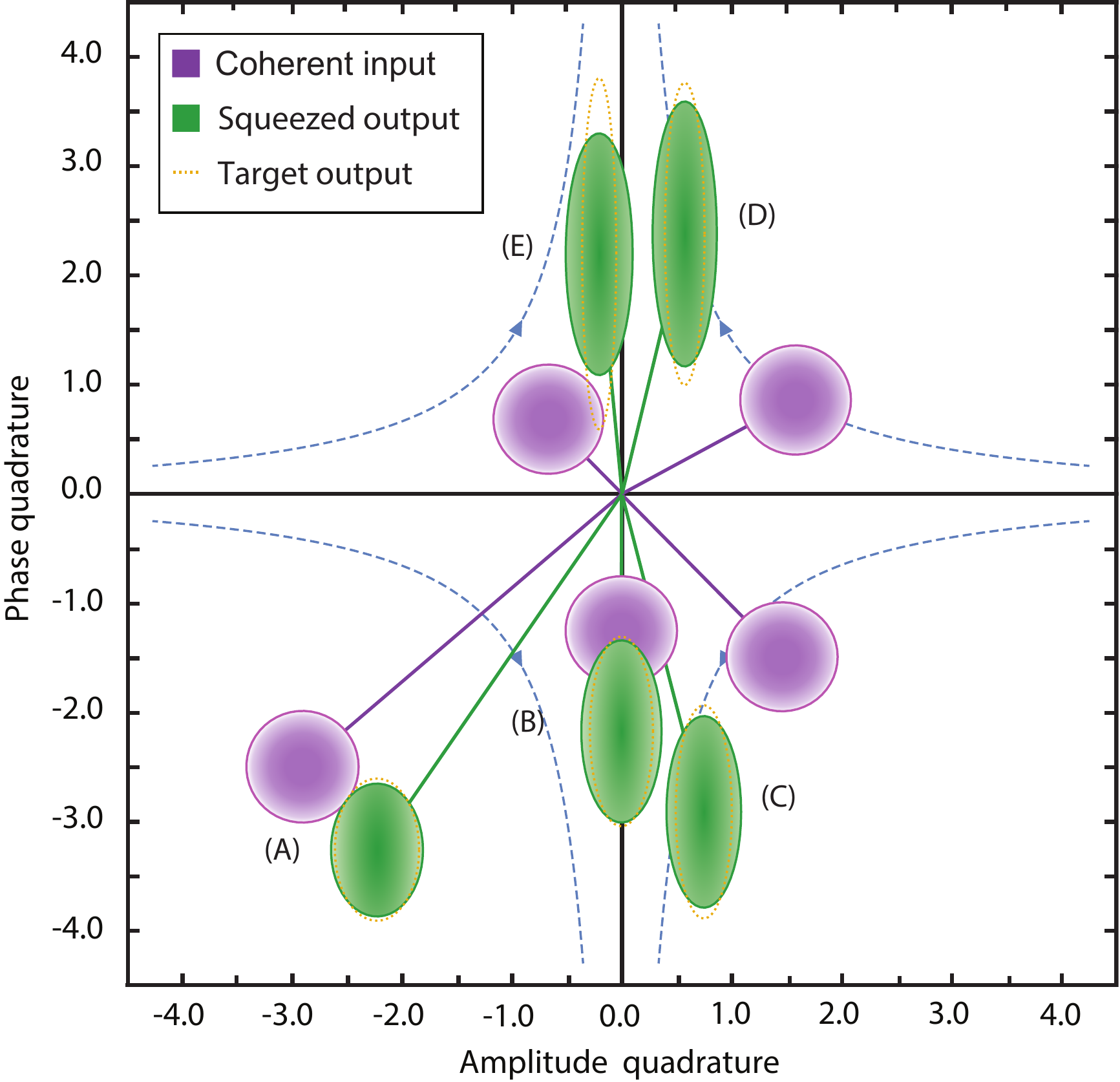}
  \caption{\textbf{Phase space diagram for the squeezing gate.} To
    verify the phase invariance of the squeezing operation,
    five coherent states, labelled (A)--(E), located at different regions of phase space are chosen as inputs. The target squeezings are \SI{2.30}{\dB}, \SI{4.81}{\dB}, \SI{5.84}{\dB}, \SI{8.85}{\dB}, and \SI{10.16}{\dB}, respectively. The corresponding inputs, the experimental and desired squeezed outputs are represented by the noise contours (one standard deviation width) of their Wigner functions. In all circumstances, the squeezing gate behaves consistently irrespective of the input amplitude or phase. }
\label{fig:2}
\end{figure}

\begin{figure}[h]
  \centering
  \includegraphics[width=1\columnwidth]{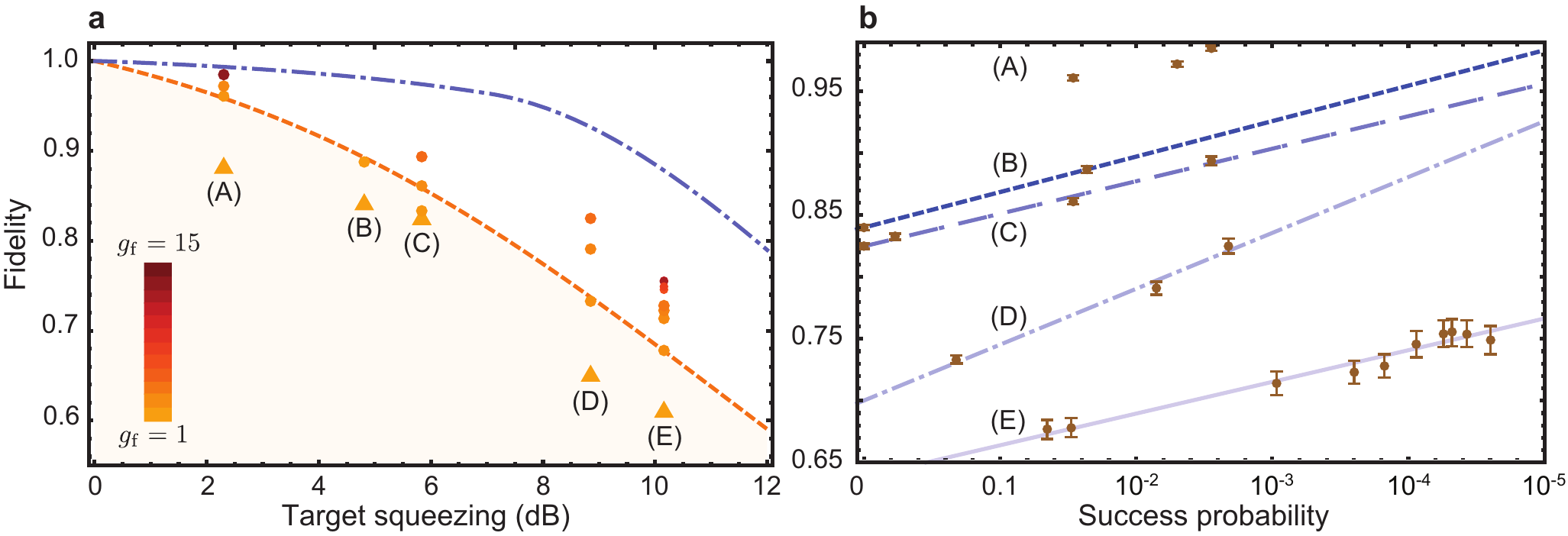}
  \caption{\textbf{Improvement in fidelity over conventional
      techniques for a series of target squeezing for states
      (A)--(E).}  \textbf{a.} The optimal fidelities attainable in two scenarios
    are plotted as performance benchmarks: the presented squeezing
    gate when a dual homodyne is performed (top blue dash-dotted curve),
    and a conventional squeezing gate (bottom orange curve). Both lines assume no experimental imperfections,
    representing the optimal fidelity attainable from our initial squeezed
    resource. Experimental results
    with target squeezing between \SI{2.30}{\dB} and \SI{10.16}{\dB} are plotted
    with round markers, showing an increase in fidelity as
    the filter strength increases (darker gradient colour). The triangle
    markers denote the fidelity obtained when filter strength is set to one. \textbf{b.} The improvement in fidelity comes at the expense of decreasing success probability. Error bars represent 1 s.d. of the output fidelity (see Supplementary section~\RNum{3}).}
\label{fig:3}
\end{figure}

\begin{figure}[t!]
\includegraphics[width = 1\columnwidth]{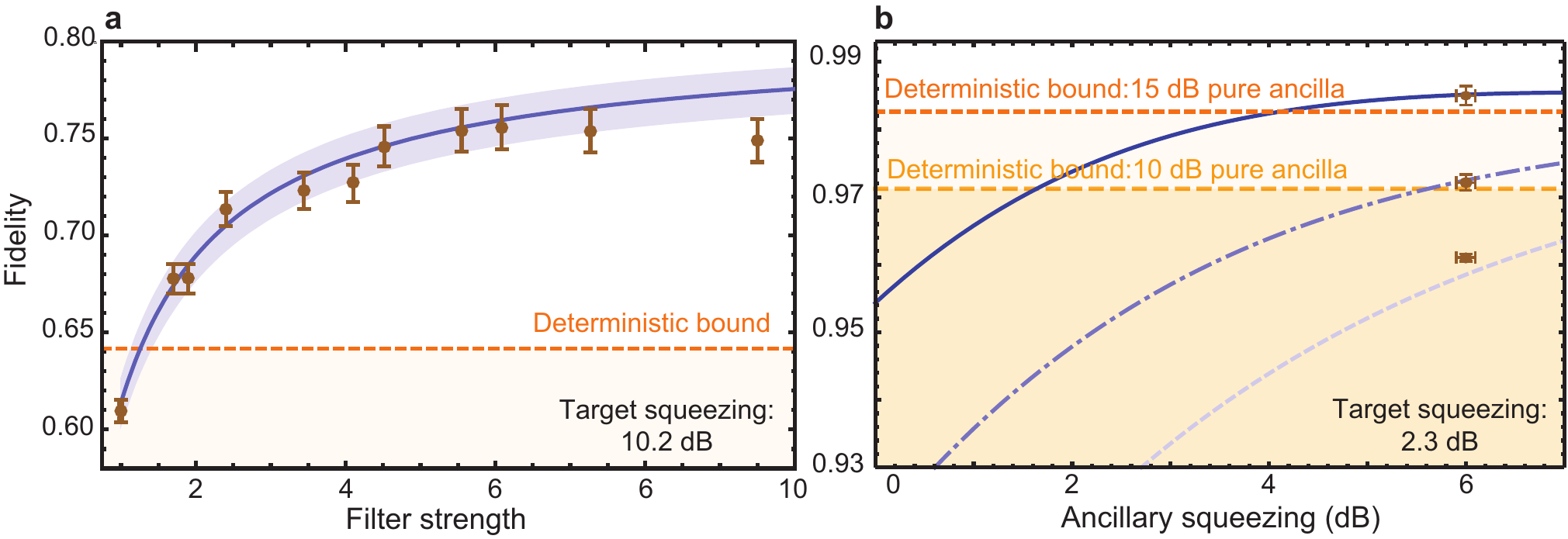}
\caption{\textbf{Fidelity as a function of filter strength and
    ancillary squeezing.} \textbf{a.} The fidelity consistently increases with
  a higher filter strength, which agrees with the theoretical
  prediction taking into account of experimental imperfections (blue
  shaded region). The orange dashed line denotes the limit of fidelity
  for deterministic protocols assuming pure \SI{6}{\dB} squeezed
  ancilla and perfect experimental conditions. Reaching fidelity
  beyond it would require a larger amount of ancillary squeezing
  conventionally. \textbf{b.} Filter strength can compensate ancillary
  squeezing strength. The requirement for a more squeezed ancilla can
  be circumvented with the presented squeezing gate. The three purple
  lines, from bottom to top, represent filter strengths of $g_\text{f}=1.52$, $3.38$ and
  $12.63$. For comparison,
  we present the achievable fidelity for a conventional squeezing gate
  operating with the currently-best-available squeezing \SI{15}{\dB}
  (orange line) and \SI{10}{\dB} squeezing (yellow line), but
  sustaining the same in-loop detection efficiency as present in our
  setup. With \SI{6}{\dB} of initial squeezing, we observe fidelity
  surpassing both bounds. Error bars represent 1 s.d. of the output fidelity and the ancillary squeezing.}
\label{fig:4}
\end{figure}

\end{document}